\documentclass[aps,pre,preprint,onecolumn,showpacs,amsmath,amssymb]{revtex4-1}
\usepackage[utf8]{inputenc}

\usepackage{graphicx}
\usepackage{bm}
\usepackage{color}
\usepackage{soul} 
\usepackage{lineno}
\usepackage{textgreek}
\usepackage{physics}

\usepackage{comment}

\begin{document}

\title{Direct observations of rising oil droplets deformed by hydrocarbonoclastic bacteria}

\author{Vincent Hickl$^{a}$}
\author{Hima Hrithik Pamu$^{b}$}
\author{Gabriel Juarez$^{b}$}
\thanks{Email address}\email{gjuarez@illinois.edu}

\affiliation{$^{a}$Department of Physics, University of Illinois at Urbana-Champaign, Urbana, Illinois 61801, USA}

\affiliation{$^{b}$Department of Mechanical Science and Engineering, University of 
Illinois at Urbana-Champaign, Urbana, Illinois 61801, USA}

\date{\today}


\begin{abstract}

In marine environments, microscopic droplets of spilled oil can be transported over tens or hundreds kilometers in the water column. 
As this oil is biodegraded, growing bacteria on the droplets' surface can deform the oil-water interface to generate complex shapes and significantly enlarge droplets. 
A complete understanding of the fate and transport of spilled oil requires bridging the present gap between these length scales and determining how microscale processes affect large-scale transport of oil. 
Here, we describe experimental results describing rising oil droplets in a purpose-built hydrodynamic treadmill which rotates to keep droplets stationary in the lab frame for continuous, direct observation.
Droplets of radii $10-100$ \textmu m are colonized and deformed by bacteria over several days before their rising speeds through the water column are measured. 
Rising speeds of deformed droplets are significantly slower as a result of bio-aggregate formation at the droplet surface compared to those of droplets of weathered or unweathered oil without bacteria.
Additionally, we discover bio-aggregate particles of oil and bacterial biofilms which sink through the water column. 
The composition of these particles is quantified using fluorescence microscopy. 
These results have important implications for the study of oil transport following a spill, as colonization by bacteria can cause oil droplets to remain in the water column for months or years longer than otherwise  expected. 
An improved understanding of the physical transformations of oil droplets during biodegradation shows promise for significantly improving models of oil spills. 
Additionally, the formation of sinking particles of oil and bacteria presents a new vector for hydrocarbon sedimentation and potential marine oil snow formation.\\

\textbf{Keywords:} biodegradation, oil-degrading bacteria, oil droplets, hydrodynamic treadmill, hydrocarbon transport, oil spills.\\

\textbf{Synopsis}: Microscopic processes affect the transport of oil droplets over long distances. This study uses novel experimental protocols to show how bacterial deformations reduce droplet rising speeds and cause oil sedimentation.

\end{abstract}

\maketitle

\section{Introduction}

Understanding the transport of crude oil through marine environments is crucial for the study of the long-term effects of anthropogenic oil spills and of natural marine oil seeps. Oil spills in particular can cause catastrophic damage to marine ecosystems, and oil spill response techniques can only be effective if the complex physics which underlie the motion of oil through the water column are considered. An important challenge is to accurately predict the transport of microscopic, immiscible droplets of oil which can be suspended at depth for months after a spill \cite{Camilli2010,Passow2021}. These droplets, which typically range from $\leq 1$ \textmu m to hundreds of \textmu m in radius \cite{Li2015}, can form plumes which contain as much as 70\% of the oil spilled in events such as the Deepwater Horizon (DWH) disaster \cite{Ryerson2012,Passow2021}. In situ observations and simulations of oil spills have shown that these droplets can be transported over hundreds of kilometers \cite{North2015,French-McCay2019}. To fully describe the fate of these droplets, one must bridge the gap between microscale processes underlying the evolution of the droplets themselves and macroscale processes determining their long-term transport.


As oil is transported through the water column, marine bacteria attach to the surface of droplets and can degrade some components of the oil. Some estimates suggest that, following the DWH disaster, 30\% of the spilled oil may have been biodegraded by marine organisms, especially bacteria \cite{Passow2016}. Accordingly, the ultimate fate of spilled oil cannot be accurately assessed without considering the effects of biodegradation on droplet transport.

Existing models of oil biodegradation and transport typically make a number of simplifying assumptions regarding droplet evolution. For example, biodegradation is taken to involve a simple shrinkage of oil droplets over time. Some authors also explicitly assume that oil droplets remain spherical, and any biofilm formation on their surface is neglected \cite{Vilcaez2013,North2015}. Additionally, it is commonly assumed that droplet buoyancy remains constant and the rate of degradation is the same for all droplets \cite{North2015}. In reality, droplet degradation rates depend on droplet size not only because of different area-to-volume ratios, but also because encounters between cells and droplets can be a rate-limiting step \cite{Fernandez2022}. Accurate predictions of sub-surface oil transport require an improved and more detailed understanding of biodegradation mechanics \cite{Vilcaez2013,North2013,Daly2016,Socolofsky2019}.

Recent work has shown that the growth of bacteria confined to the surface of oil droplets can cause deformations of the interface \cite{Hickl2022}. As cells continue to grow, an aggregate of biofilm and oil forms around droplets, enlarging them by as much as a factor of 2 \cite{Omarova2019,Hickl2022b}. This observation contradicts the prevailing notion that droplets must shrink as they are degraded. Understanding how droplet deformations affect the transport of oil is crucial for determining its eventual fate. By pinning disks of oil between two surfaces, it has been shown that interfacial biofilms increase the drag on oil droplets \cite{White2020}. However, the magnitude of this change, or its dependence on droplet size and morphology remain unknown. Additionally, other effects such as changes in buoyancy have yet to be described. To our knowledge, there are no experimental observations of how droplets colonized and deformed by oil-degrading bacteria are transported vertically through a water column. 

The vertical transport of hydrocarbons is a crucial ecological process which can affect the global climate by reducing atmospheric $\text{CO}_2$ via sedimentation in the ocean \cite{Kiorboe2001}. Following oil spills, most of the oil either rises to form surface slicks or becomes entrenched in deep-sea plumes. In both cases, marine snow particles can cause hydrocarbons to sink and become sedimented on the ocean floor. Estimates have suggested that as much as 14\% of oil spilled in the Deepwater Horizon Disaster may have sunk to the seafloor, largely as a result of marine snow formation \cite{Daly2016,Valentine2014}. A variety of mechanisms for the formation of marine snow in the context of oil spills have been described \cite{Brakstad2018,Ross2021}, but most emphasize the importance of phytoplankton which interact with oil and other organisms to form particles $>0.5$ mm in size \cite{Gregson2021}. Less attention has been paid to smaller sinking particles which may include spilled oil. In particular, while bacteria are known to attach to existing marine snow \cite{Kiorboe2001,Slomka2020}, it is unclear to what extent their growth can itself lead to the formation of sinking, hydrocarbon-rich particles. Recent observations of biofilm formation on both oil films \cite{Omarova2019} and droplets \cite{Abbasi2018,Hickl2022}, suggest that oil dispersion caused by bacterial colonization may be important to consider.

Here, we present experimental results from direct observations of the vertical transport of oil droplets colonized and deformed by bacteria. Inspired by recent advances in the study of swimming microorganisms~\cite{Krishnamurthy}, we construct a ``hydrodynamic treadmill," which allows for the stationary observation of droplets (in the lab frame), while maintaining continuous motion through a water column. Using this method, precise measurements of oil rising and sinking speeds are achieved in vitro. Oil droplets colonized by bacteria are deformed significantly, which can lead to the formation of aggregates of oil and biofilm with complex morphologies. Droplets deformed by bacteria rise significantly slower than droplets of both unweathered and weathered oil as a result of changes in both drag and buoyancy. As a result of oil dispersion by bacterial colonization, bio-aggregates of oil and biofilm sink through the water column. The composition of these aggregates varies widely, with crude oil comprising between $0.05\%$ and $7.8\%$ of the volume. These results challenge several key assumptions in the literature on the transport of suspended oil subject to biodegradation. The sinking of aggregates of oil and bacteria represents a new pathway through which microscopic, hydrocarbon-rich particles can sediment in the ocean.

\section{Experimental methods}

To visualize rising oil droplets, we constructed a rotating annular fluid chamber which acts as a hydrodynamic treadmill for the droplets. The chamber was constructed using aluminum rings (3 mm thick) sandwiched between laser cut sheets of acrylic (1.5 mm thick). The inner aluminum ring had an inner radius of 15 mm and an outer radius of 45 mm. The outer ring had an inner radius of 65 mm and an outer radius of 80 mm. The four pieces were screwed together and sealed using silicone rubber sheets (1 mm thick) and epoxy. Thus, the fluid chamber was a ring of width 20 mm, with a radius of 55 mm at the center, and a thickness of 4 mm. The chamber was mounted on a 8 mm steel shaft (Phidgets) using a universal mounting hub (Pololu).

The chamber was rotated by a DC motor (McMaster-Carr) via a custom-built gearbox with laser-cut acrylic gears to reduce the speed of rotation and provide finer control over the angular velocity. The speed of the motor could be set manually using a power controller, or programmatically using a motor driver carrier (Pololu VNH5019) and an Arduino microcontroller. The angular speed of the chamber could be controlled with a resolution of less than $5\times10^{-4}$ rad/s. An optical encoder (Phidgets IHC3808) was used to measure the precise angular velocity of the chamber during experiments. These components were mounted onto an optical breadboard using optical posts, rotary bearings, and various custom 3D printed and laser cut parts.

Images were taken at $2.3 \times$ magnification using a Ximea SCMOS camera and a zoom lens, mounted on a tripod and focused on the 3 o'clock position of the annular fluid chamber. 
Each droplet is recorded for 10 s at 20 fps. 
Images are calibrated in each experiment using a ruler placed in the field of view. The resolution is $2.37$ \textmu m/pixel.

\begin{figure*}
    \centering
    \includegraphics[width=\linewidth]{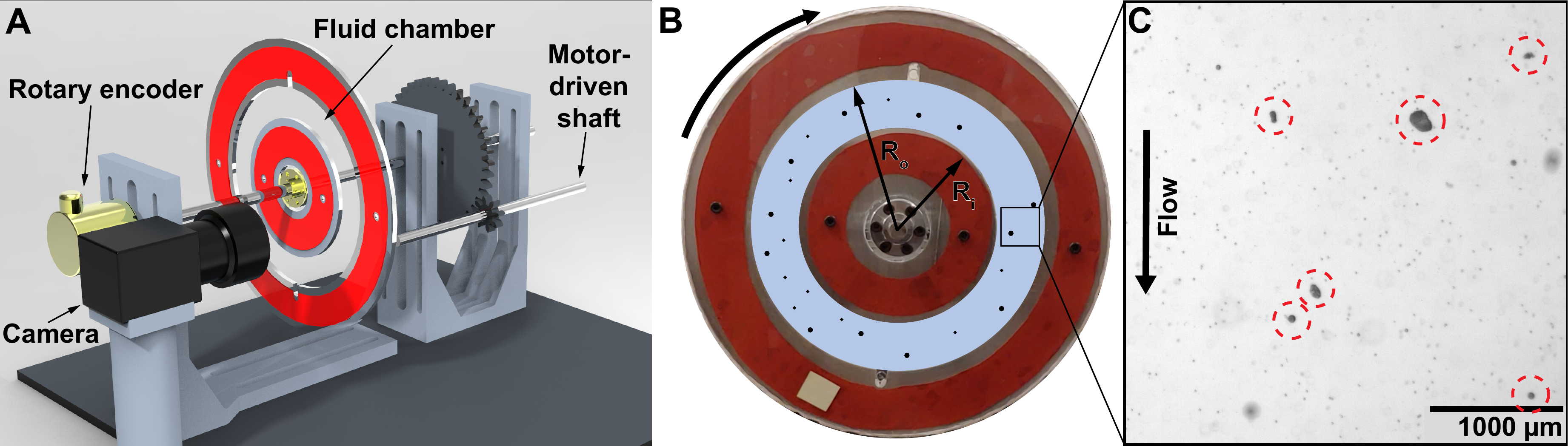}
\caption{Hydrodynamic treadmill. 
\textbf{(A)} Render of our setup assembly, showing the fluid chamber, custom gearbox, camera, and optical encoder mounted on an optical breadboard. 
\textbf{(B)} Frontal view of annular fluid chamber with inner and outer radii $R_i$ and $R_o$, with camera fixed on the 3 o’clock position. 
\textbf{(C)} Sample image of deformed oil droplets (red circles) in tracer particle suspension in the fluid chamber.
} 
    \label{fig:fig1}
\end{figure*}

The bacterial strain used in this study was \emph{A. borkumensis} (ATCC 700651), a rod-shaped, non-motile, alkane-degrading marine bacterium. Bacteria were grown by incubating cells in 5 mL of culture medium for $24$ hours at $30 \ ^{\circ}$C in an orbital shaker at $180$ rpm. The culture medium consisted of 37.4 g/L of 2216 marine broth (BD Difco) and 10 g/L of sodium pyruvate. The average doubling time was measured to be 1.6 hours using a Biowave CO8000 cell density meter.

In each experiment $0.5$ mL of MC525 crude oil was added to a culture tube containing culture medium inoculated with bacteria, forming a thin oil film on the water surface. MC252 is a light sweet crude oil having a density of $0.850$ g/mL, an interfacial tension of $20$ mN/m, and a viscosity of $3.9$ mPa s at $32 \ ^{\circ}$C \cite{Daling2014}. This oil is autofluorescent throughout the visible spectrum. The tube containing cell culture and oil was then incubated for $3-5$ days at $30 \ ^{\circ}$C in an orbital shaker at $180$ rpm. The agitation produced a polydispersion that includes oil droplets with radii $10$ \textmu m $< R < 100$ \textmu m to which bacteria can attach, grow, and induce deformations.

There were two control protocols: one with weathered and one with unweathered oil. For weathered oil, the sample preparation was identical to that described above, except no cells were added to the culture medium. This protocol enables us to control for any evaporation or dissolution of oil that occurs while bacteria grow. For unweathered oil, a polydispersion of oil in DI water is produced via gentle agitation in a test tube, without any incubation. This protocol is used to confirm that the experimental setup and measurement protocol is working as intended, since the density and expected rising speeds of the unweathered oil are known. 

In all three protocols (deformed oil, unweathered oil, and weathered oil), the polydispersion of oil in water was injected into the rotating fluid chamber. To accurately measure the rising speeds of droplets relative to the surrounding fluid, polystyrene tracer particles with a $16.6$ \textmu m diameter were added to the fluid at a concentration of $8\times10^3$ particles/mL. In videos of droplets suspended in the rotating chamber, the fluid speed $v_f$ was determined with a custom PTV MATLAB algorithm used to determine the velocity of the tracer particles surrounding each droplet. The velocity of the droplets relative to the camera $v_{oc}$ was similarly determined. The velocity of the droplets relative to the surrounding fluid is then given by $v_{of}=v_{oc}-v_f$.

\section{Results}

\begin{figure*}
    \centering
    \includegraphics[width=\linewidth]{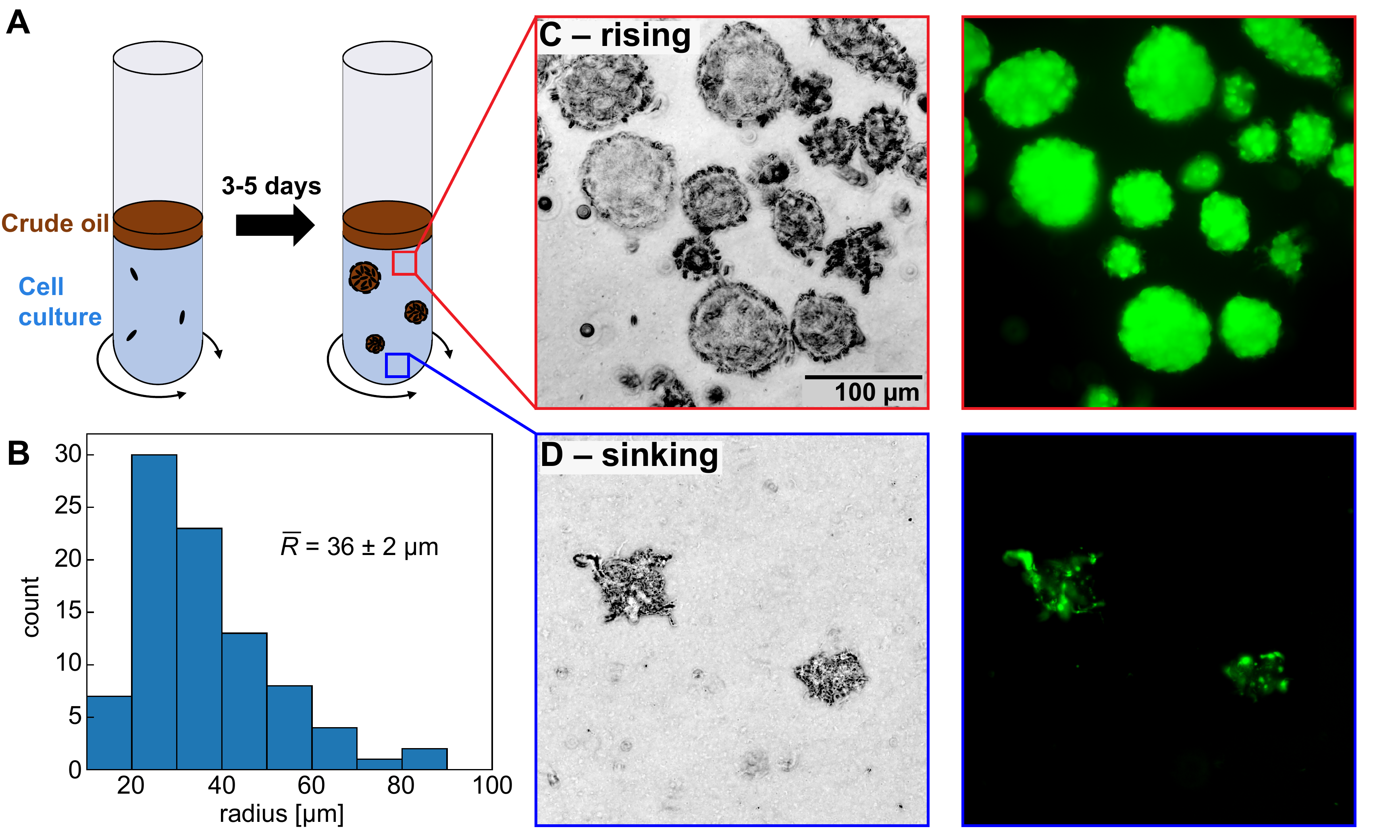}
\caption{Droplet formation and morphology. 
\textbf{(A)} Schematic of the protocol for polydispersion of oil droplets into a cell suspension.
Bacteria are incubated for several days in a liquid culture with a film of oil at the surface. 
Constant agitation disperses oil into the fluid, where droplets are colonized and deformed. 
\textbf{(B)} Droplet size distribution for experiments with and without bacteria. 
\textbf{(C)} Deformed oil droplets with positive buoyancy. 
Phase contrast (left) and fluorescence (right) microscopy images show oil droplets buckled and tubulated by bacteria. 
\textbf{(D)} Aggregates of oil and biomass with negative buoyancy. 
Phase contrast images (left) show full size of sinking object. 
Fluorescent images (right) show that only a fraction of the aggregate is crude oil. Scale bar is the same for (C) and (D).
} 
    \label{fig:fig2}
\end{figure*}

Droplets suspended in a cell suspension become colonized and deformed by bacteria. Suspended droplets are formed by continuously agitating tubes containing a cell culture with a thin layer of oil, as shown in Fig. \ref{fig:fig2}A. Then, bacteria attach to and grow at the droplet surface, and eventually overcome surface tension to deform droplets. After several days, a small amount of the oil-in-water dispersion is taken out of the tube and placed between a microscope slide and cover slip separated by 1 mm spacers. Phase contrast and fluorescence microscopy are then used to visualize the suspended oil particles along with any attached bacteria and biofilm. Nearly all droplets appear buckled, with many having tube-like protrusions at their surface, as shown in Fig. \ref{fig:fig2}C-D. The radii of all droplets from experiments in the hydrodynamic treadmill (including controls) are shown in Fig. \ref{fig:fig2}B. For deformed droplets, the equivalent radius is given by $R_{eq}=\sqrt{A_c/\pi}$, where $A_c$ is the cross-sectional area of the droplet in the imaging plane. The mean droplet size measured in the treadmill was $\Bar{R}=36\pm2$ \textmu m.

The composition and morphology of oil particles determine their buoyancy. Once injected onto the microscope slide, some aggregates of oil and biofilm rise to the cover slip, while others sink and remain on the glass slide. Since the oil is autofluorescent, fluorescence microscopy was used to determine how the oil was distributed within particles with both positive and negative buoyancy. Rising particles appear to be buckled droplets that consist primarily of oil, with some bacteria and biofilm at the interface, as shown in Fig. \ref{fig:fig2}C. On the other hand, sinking particles consist primarily of bacteria and biofilm, which have smaller oil particles suspended within them, as shown in Fig. \ref{fig:fig2}D.

\begin{figure*}
    \centering
    \includegraphics[width=0.8\linewidth]{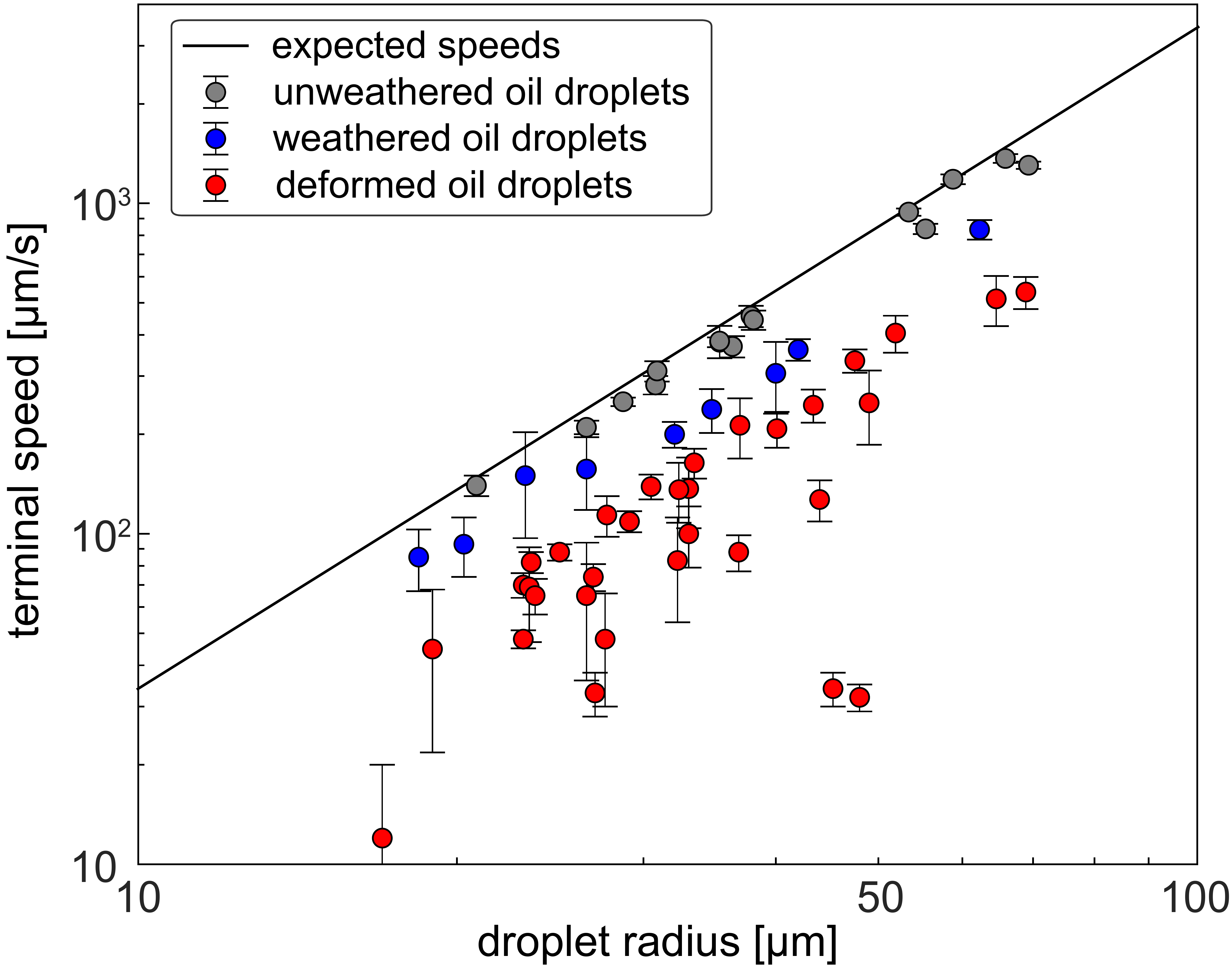}
\caption{Measurements of droplet rising speeds from hydrodynamic treadmill. 
(Black line) The expected rising speeds for unweathered droplets with no bacteria as determined from Eq. 1.
(Grey and Blue points) Experimentally measured rising speeds of unweathered and weathered oil droplets, respectively, that are not colonized or deformed by bacteria. 
(Red points) experimentally measured rising speeds of deformed droplets that are visibly non-spherical as a result of cell attachment and growth. 
Error bars represent variation in speed from PTV measurements. 
} 
    \label{fig:fig3}
\end{figure*}

The hydrodynamic treadmill gives accurate measurements of the rising speeds of uncolonized and unweathered oil droplets. In control experiments, a polydispersion of oil in water is injected into the rotating chamber. For microscopic droplets, the Reynold's number $Re=\rho_w v_TR/\mu<<1$, meaning viscous forced dominate over inertial ones. Here, $v_T$ is the terminal speed, and $\mu$ and $\rho_w$ are the viscosity and mass density of water, respectively. Therefore, the expected terminal rising speed of a droplet is calculated by setting the sum of the weight, buoyant force, and Stokes' drag equal to zero. For spherical droplets, this gives 
\begin{equation}
    v_T=\frac{2}{9}\frac{g}{\mu}(\rho_o-\rho_w)R^2
\end{equation}
where $g$ is the acceleration due to gravity, and $\rho_o$ is the density of the oil. 
All droplets analyzed in the treadmill had Reynolds numbers $Re<0.1$.
Since the density of unweathered oil is known, this speed can readily be calculated for droplets of unweathered oil, shown in Fig. \ref{fig:fig3} (solid line).
Experimental measurements of rising speeds of unweathered droplets are in very good agreement with this prediction, shown in Fig. \ref{fig:fig3} (grey points). 
This result validates our method for measuring vertical speeds of microscopic objects in a water column.

Droplets of weathered oil rise more slowly than droplets of unweathered oil. 
These droplets are produced by maintaining a polydispersion of oil droplets in culture medium on an orbital shaker for 5 days. 
During this time, volatile components of the crude oil can evaporate, while soluble components are dissolved in the surrounding fluid. 
Since these components tend to be lighter \cite{Daling2014}, this weathering process is expected to increase the density of the oil droplets. 
As expected, droplets of weathered oil rise more slowly in the hydrodynamic treadmill, shown in Fig. \ref{fig:fig3} (blue points). 
Using these speeds, we estimate the density of weathered oil to be $0.899\pm0.013$ g/mL, which is significantly higher than the density of unweathered oil, $0.850$ g/mL.

Oil droplets deformed by bacteria rise significantly more slowly than both unweathered and weathered droplets of oil. 
Droplets are deformed by maintaining a polydispersion of oil in culture medium with bacteria for $3-5$ days. 
As bacteria grow on the droplet surface, they form a monolayer which produces active stresses that can overcome surface tension to deform the oil-water interface. 
Droplets than are visibly deformed when observed in the hydrodynamic treadmill all have smaller rising speeds than droplets that are only weathered without bacteria, shown in Fig. \ref{fig:fig3} (red points).
The mean ratio of the rising speed of deformed droplets to the theoretical rising speed obtained from Eq. 1 was $v_D/v_T(R)=0.46\pm0.19$, with values ranging from $0.05$ to $0.69$. 
Here, $R$ is the radius of deformed droplets measured directly in the hydrodynamic treadmill. 
The variation in speeds is much larger than in control experiments, which is due to the complex and varied morphologies of different droplets, as shown in Fig.~\ref{fig:fig2}C.

Bio-aggregates of oil and bacteria that do not rise in water were also observed using the hydrodynamic treadmill. By reversing the direction of the rotation of the fluid chamber, continuous observation of sinking particles was possible. Additionally, some particle were nearly stationary, meaning they were neutrally buoyant. Both sinking and neutrally buoyant particles have highly complex, non-spherical morphologies. As a result, precisely measuring their size was not possible in the hydrodynamic treadmill, as only a 2D cross-section could be observed. For sinking particles, the measured speeds ranged from approximately $10$ \textmu m/s to $140$ \textmu m/s. 

\begin{figure*}
    \centering
    \includegraphics[width=\linewidth]{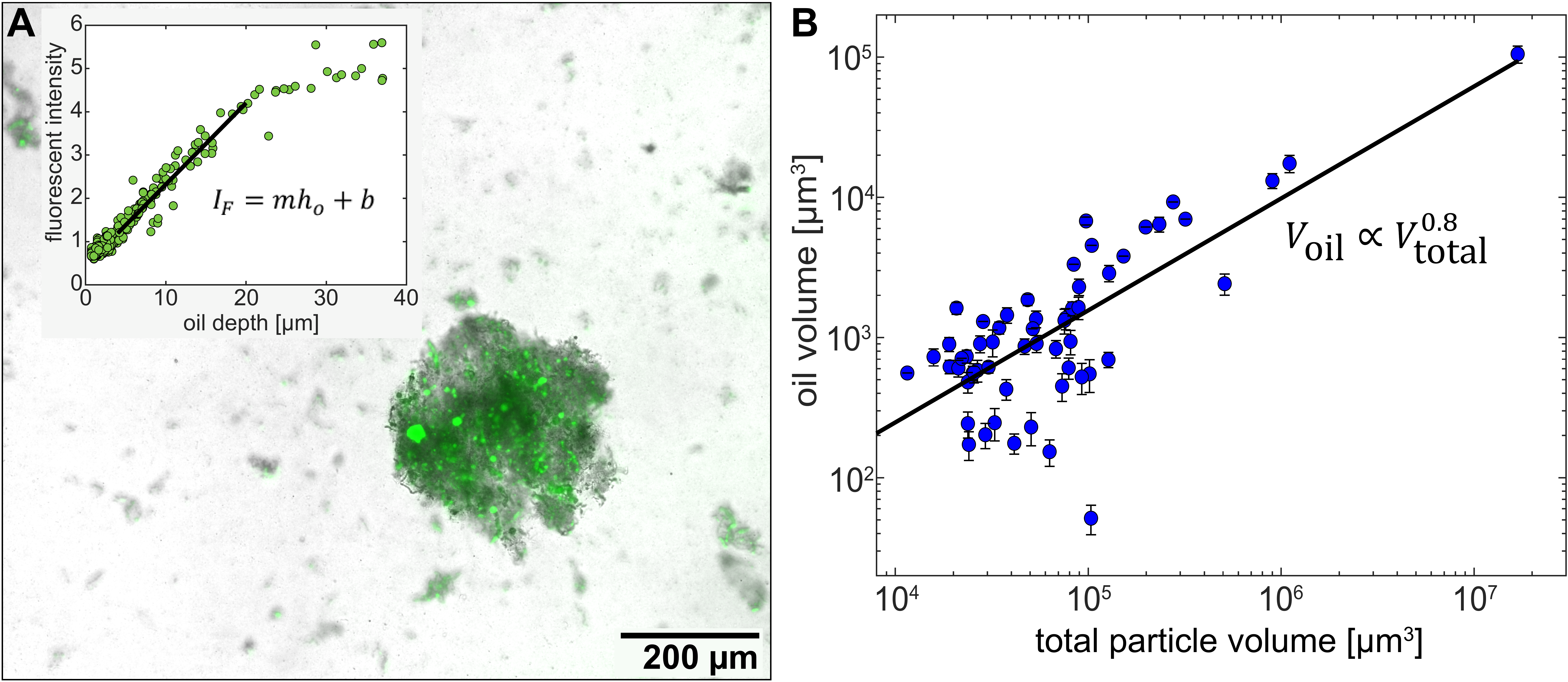}
\caption{Composition of sinking particles of oil and bacteria.
\textbf{(A)} Composite image of a sinking particle with approximate radius of $159$ \textmu m, taken using brightfield (grayscale) and fluorescence microscopy (green false color, showing crude oil) at $10\times$ magnification. 
\textbf{(Inset)} The calibration curve of fluorescence intensity as a function of oil depth used to estimate the volume of oil in such particles. 
\textbf{(B)} Log-log plot of estimated oil volume in sinking particles of different volumes. 
Error bars represent uncertainty in the quantity of oil based on the calibration curve.
} 
    \label{fig:sinking_short}
\end{figure*}

To quantify the composition of sinking particles of oil and bacteria, an oil-in-water dispersion was created by letting bacteria colonize oil in a test tube under constant agitation over 5 days, as described above. Then, the agitation was stopped, and the test tube was kept stationary for $1$ h, allowing negatively buoyant particles to sink, and positively buoyant particles to rise to the surface. 
The fluid at the bottom of the tube was carefully decanted to prevent further mixing, and a sample was placed between a microscope slide and a cover slip separated by 1 mm spacers. 
After an additional $0.5$ h of waiting, particles that had settled onto the glass slide were imaged using brightfield and fluorescence microscopy. Brightfield microscopy showed the entire particles, including oil and bacteria, while fluorescence microscopy was used to determine the exact distribution of crude oil, as shown in Fig. \ref{fig:sinking_short}A.

The volume of oil in bio-aggregates was measured using fluorescence microscopy. First, stationary droplets of crude oil were formed on a glass slide in DI water without bacteria. The intensity of the fluorescent emission at the center of each droplet was measured, with the focus set to the midplane of the droplets. Plotting this intensity as a function of droplet diameter shows how the depth of oil along the optical axis (the direction perpendicular to the focal plane) determines the intensity of the fluorescent emission. 
For droplets up to a diameter of about $20$ \textmu m, the intensity depends linearly on the oil depth, as shown in Fig. \ref{fig:sinking_short}A (inset). 
This method yields a calibration curve relating the depth of the oil $h_o$ at a point to the fluorescence intensity $I_F$: 
\begin{equation}
    I_F=mh_o+b \text{ for } h_o<20
\end{equation}
where $m$ and $b$ are fitting parameters.

Sinking particles consist mostly of bacteria and their extracellular polymeric substances (EPS), and contain only a small amount of oil. In sinking particles of bio-aggregate, oil is dispersed into many (non-spherical) microdroplets with radii $R<10$ \textmu m, as shown in Fig.~\ref{fig:sinking_short}A. Therefore, the quantity of oil at each pixel of size $\delta x$ can be calculated by:
\begin{equation}
    V_{\text{oil}}(x,y)=\frac{I_F(x,y)-b}{m}\delta x^2
\end{equation}
where $(x,y)$ are coordinate in the image plane. Then, to calculate the total volume of oil in a bio-aggregate particle, the volume of oil is summed over all pixels in the region $P$ where the particle is located:
\begin{equation}
    V_{\text{oil}}=\sum_{\{x,y\}\in P}\max\left(\frac{I_F(x,y)-b}{m},0\right)\delta x^2
\end{equation}
The $\max$ function serves to ensure that the calculated oil volume at each pixel is non-negative. 

Fitting suggests a power law relationship of $V_{\text{oil}}\propto V_{\text{total}}^{0.8}$, as shown in Fig.~\ref{fig:sinking_short}B, meaning the proportion of oil tends to decrease slightly with particle size. 
The mean percentage of oil across all particles with negative buoyancy is $2.3\pm0.2\%$, with values ranging from $0.05\%$ to $7.8\%$, meaning most of the particle volume is composed of cells and biofilm.
The low minimum value of the proportion is consistent with our observation that some sinking particles appear not to contain any oil, suggesting the biofilm formed at the oil-water interface can break apart.

\section{Discussion}

Direct observation of rising oil droplets deformed by oil-degrading bacteria in the hydrodynamic treadmill revealed that these droplets rise significantly slower through the water column than either unweathered or weathered droplets without bacteria. Additionally, sinking particles of oil and bacterial bio-aggregates were discovered. Here, we perform further analysis to rigorously quantify the effect of bacterial colonization on droplet rising speeds. Additionally, we will make quantitative estimates of the composition of sinking particles and their effect on vertical transport of oil. When measuring particle sizes in the hydrodynamic treadmill, it was not possible to distinguish between oil and bio-aggregates around droplets. In previous work using microfluidics, it was shown that bio-aggregate formation causes droplets to increase in size by some $\Delta R$ \cite{Hickl2022}. The value of $\Delta R$ varied significantly between droplets, making it difficult to accurately estimate the initial radius of deformed droplets in the hydrodynamic treadmill. 

Here, previous measurements of $\Delta R$ are used in conjunction with rising speeds measured in the hydrodynamic treadmill to estimate the initial radius $R_0$ of deformed droplets in the treadmill. Deformed droplets observed in the treadmill were sorted by how much slower their observed rising speed $v_D$ was compared to the expected speed $v_T(R)$ calculated from Eq. 1, where $R=R_0+\Delta R$ is the total radius. It is assumed that for the droplet with the largest value of $v_D/v_T(R)$, the relative size of the bio-aggregate is near its minimum measured value of $\Delta R/R_0=0.08$. Conversely, it is assumed that for the deformed droplet with the smallest value of $v_D/v_T(R)$, the relative size of the bio-aggregate is at its maximum value of $\Delta R/R_0=1.56$. By extrapolating to the rest of the droplets, the initial radius $R_0$ of each droplet in the hydrodynamic treadmill can be estimated (see Supplemental Information for more details). Then, the relative rising speeds of deformed droplets were calculated to be $v_D/v_T(R_0)=0.72\pm0.15$, on average, with values ranging from $0.30$ to $0.86$. As expected, the rising speeds of small droplets tend to change the most, since the relative increase in size $\Delta R/R_0$ is inversely correlated with the initial radius $R_0$ \cite{Hickl2022}.

\begin{figure*}
    \centering
    \includegraphics[width=0.8\linewidth]{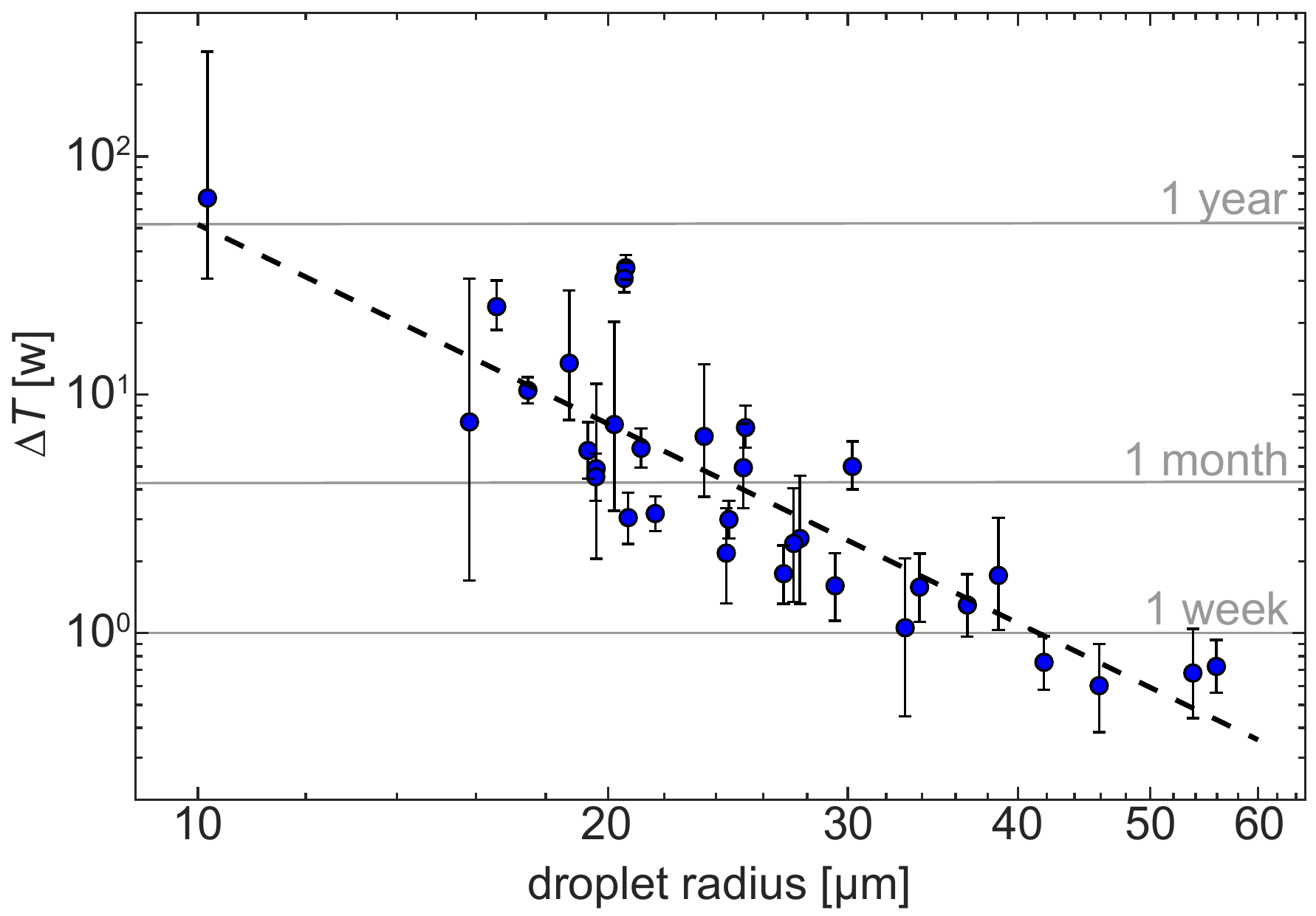}
\caption{Change in rising time due to droplet deformation from cell growth. 
(Blue points) The estimated increase in the rising time through the water column from a depth of 1000 m as a result of droplet deformation from bacterial colonization.
For reference, horizontal lines indicate 1 week, 1 month, and 1 year. 
Droplet radii are estimated from previous results describing bio-aggregate emergence due to interfacial cell growth. 
The dashed line represents best power-law fit to the data: $\Delta T\propto R_0^{-2.8}$.
} 
    \label{fig:rising_change}
\end{figure*}


Based on these calculations, the effect of droplet deformations by interfacial cell growth on the long range transport of spilled oil can be estimated. The time it takes for droplets of weathered oil without bacteria to rise to the ocean surface from a typical deep-sea spill at 1000 m is calculated using the terminal speed from Eq. 1. Values of this expected rising time for the initial droplet size range estimated from our experiments, $10\leq R_0\leq60$ \textmu m, range from 2 weeks for $60$ \textmu m droplets to about a year for $10$ \textmu m droplets. The change in the rising time $\Delta T$ resulting from bacterial colonization and deformation can then be estimated for the experimentally observed droplets, as shown in Fig.~\ref{fig:rising_change} (blue points). For the largest droplets with $R\gtrsim50$ \textmu m, the change in rising time is several days, a significant increase to the time this oil remains in the water column. For the smallest droplets with $R\approx10$ \textmu m, the rising time can change dramatically, increasing by 1-2 years. The changes in rising time $\Delta T$ appear to depend on initial droplet radius following a power law $\Delta T\propto R_0^\alpha$, with $\alpha=-2.8\pm0.3$, as shown in Fig.~\ref{fig:rising_change} (dashed line). 

These calculations suggest that the additional time that droplets remain in the water column is sensitive to the amount of bio-aggregate formed at the droplet surface. It is important to consider the change in droplet size due to bacterial colonization when estimating the change in rising times (see Supplemental Information for further discussion). In reality, droplets in this size range became suspended in a deep-sea plume at around 1200 m for months following the DWH spill~\cite{Camilli2010,Hazen2010}. It is possible that colonization by bacteria contributed to the persistence of these droplets at depth. Alternatively, the emergence of a bio-aggregate which decreases the buoyancy of the droplets could cause them to sink from the plume, contributing to the known phenomenon of oil sedimentation \cite{Daly2016}. More work is needed to investigate the transport of droplets under different flow conditions that mimic the environment of deep-sea plumes.

While the rising particles observed in the above experiments were droplets colonized and deformed by cells, sinking particles had more complex morphologies. They were large pieces of bio-aggregate in which small quantities of oil were embedded. There are two primary mechanisms by which such aggregates could be formed in the setup described in Fig.~\ref{fig:fig2}. First, if a large bio-aggregate forms around the droplet, it is possible that parts of it can break off because of the agitation of the dispersion. Second, a bio-aggregate could also form on the thin film of oil which remains on the culture medium as the dispersion is formed. Pieces of this aggregate could then also break off and be dispersed into the fluid. The formation of a biofilm on the underside of an oil film has previously been observed~\cite{Omarova2019}. The distinction between these two formation mechanisms of sinking bio-aggregates are important, as they describe how bacterial colonization can lead to downward transport of oil both from a deep-sea plume of droplets and from an oil slick at the ocean surface.

The total quantity of oil transported downwards by sinking particles of bio-aggregate is very small. By using fluorescence microscopy to measure the volume of oil in many sinking particles, the density of oil in the fluid from which samples were taken is estimated to be between 1.6 and 8.1 ppm. The volume of oil initially used in these experiments was $100$ \textmu L, and the total sample volume for sinking particles was $2$ mL. Therefore, only between $3.2\times10^{-3}$ \% and $1.6\times10^{-2}$ \% of the oil became sedimented as a result of bio-aggregate formation. This particular result is likely very sensitive to the initial conditions of our experiment, particularly the amount of agitation when creating the dispersion, and the thickness of the oil slick in the tube, shown in Fig.~\ref{fig:fig2}A. Future work should consider using parameters realistic for real oil spill scenarios, potentially including the use of chemical dispersants, to more accurately assess the potential for oil sedimentation by bacterial growth.

The speeds of sinking particles measured in the treadmill, along with particle compositions measured using microscopy, can be used to estimate the mass density of biofilms of \textit{A. borkumensis}. Consider the slowest sinking particles (or neutrally buoyant particles), which have a density of $1.0$ g/mL. The proportion of oil in these particles is likely close to the maximum value observed for sinking bio-aggregates, $7.8\%$ by volume. Since the density of the oil ($0.899$ g/mL) is known, the density of the remaining volume, composed of bacteria and EPS, must be at least $1.009$ g/mL. On the other hand, consider the fastest sinking particle observed in the hydrodynamic treadmill, with a terminal speed of $142$ \textmu m/s. Supposing this particle had an average oil proportion, $2.3\%$, Eq. 1 can be used to calculate that the density of the remaining constituents of the particle must be $1.037$ g/mL. Thus, the density of the cells and EPS which make up most of the bio-aggregates is only slightly larger than that of water. This result is in agreement with other studies which have reported slowly sinking millimeter-sized particles consisting largely of exopolymers \cite{Bochdansky2016}. These parameters could be crucial for the creation of more accurate models of oil transport during bacterial colonization.


The experimental setup described here is a major step forward in the study of the transport of spilled oil. A major challenge in marine microbiology is the design of experimental techniques to investigate how microscale interactions affect macroscopic ecological processes~\cite{Taute2020}. Microorganisms as small as hundreds of micrometers in size have been shown to swim over distances of tens of meters~\cite{Blasco1978}, making direct long-term observations of individual organisms in vitro impossible. Very recently, vertical tracking microscopy techniques have been developed to study the swimming of individual plankton~\cite{Krishnamurthy}. Here, a similar hydrodynamic treadmill was used for the continuous direct observation of suspended oil droplets deformed by bacteria. The rotation of the fluid chamber allows particles of interest to be maintained in place relative to the imaging apparatus while effectively being transported indefinitely through a water column. Then, direct observation allowed for the precise characterization of how the physical forces acting on droplets in marine environments are affected by microbial growth at the interface. 

The importance of the transport of living matter in marine environments extends far beyond the subject of oil spills. Bacteria are key components of marine food webs, and the transport of hydrocarbons is crucial for the carbon cycle that regulates oceanic life \cite{DeVries2022,Moran2022}. The formation of aggregates of organic and inorganic materials, especially in the form of marine snow, is an important process that allows for nutrients to be transported into lower layers of the ocean~\cite{Kiorboe2001}. The formation of marine snow begins when particles with sizes on the order of nanometers or micrometers encounter dead or dying phytoplankton, protists, or animals. Other studies have focused on the ability of bacteria to encounter and attach to existing sinking particles~\cite{Datta2016,Nguyen2022}. Our results show that the reverse is possible as well: bacteria can form bio-aggregates by themselves, even when no other organisms or large particles are present in the water column. Other organisms can then attach to these aggregates, further promoting sedimentation of particulate organic matter. The experimental techniques described here can be adapted to further study the interactions between oil droplets and other marine organisms, and could significantly improve our understanding of the formation and transport of marine oil snow.

\section{Acknowledgments}

This material is based on work supported by the Oil Spill Recovery Institute under Contract No. 20-10-06. 

\bibliographystyle{vancouver}
\bibliography{bibliography}

\begin{thebibliography}{33}%
\makeatletter
\providecommand \@ifxundefined [1]{%
 \@ifx{#1\undefined}
}%
\providecommand \@ifnum [1]{%
 \ifnum #1\expandafter \@firstoftwo
 \else \expandafter \@secondoftwo
 \fi
}%
\providecommand \@ifx [1]{%
 \ifx #1\expandafter \@firstoftwo
 \else \expandafter \@secondoftwo
 \fi
}%
\providecommand \natexlab [1]{#1}%
\providecommand \enquote  [1]{``#1''}%
\providecommand \bibnamefont  [1]{#1}%
\providecommand \bibfnamefont [1]{#1}%
\providecommand \citenamefont [1]{#1}%
\providecommand \href@noop [0]{\@secondoftwo}%
\providecommand \href [0]{\begingroup \@sanitize@url \@href}%
\providecommand \@href[1]{\@@startlink{#1}\@@href}%
\providecommand \@@href[1]{\endgroup#1\@@endlink}%
\providecommand \@sanitize@url [0]{\catcode `\\12\catcode `\$12\catcode
  `\&12\catcode `\#12\catcode `\^12\catcode `\_12\catcode `\%12\relax}%
\providecommand \@@startlink[1]{}%
\providecommand \@@endlink[0]{}%
\providecommand \url  [0]{\begingroup\@sanitize@url \@url }%
\providecommand \@url [1]{\endgroup\@href {#1}{\urlprefix }}%
\providecommand \urlprefix  [0]{URL }%
\providecommand \Eprint [0]{\href }%
\providecommand \doibase [0]{http://dx.doi.org/}%
\providecommand \selectlanguage [0]{\@gobble}%
\providecommand \bibinfo  [0]{\@secondoftwo}%
\providecommand \bibfield  [0]{\@secondoftwo}%
\providecommand \translation [1]{[#1]}%
\providecommand \BibitemOpen [0]{}%
\providecommand \bibitemStop [0]{}%
\providecommand \bibitemNoStop [0]{.\EOS\space}%
\providecommand \EOS [0]{\spacefactor3000\relax}%
\providecommand \BibitemShut  [1]{\csname bibitem#1\endcsname}%
\let\auto@bib@innerbib\@empty
\bibitem [{\citenamefont {Camilli}\ \emph {et~al.}(2010)\citenamefont
  {Camilli}, \citenamefont {Reddy}, \citenamefont {Yoerger}, \citenamefont
  {{Van Mooy}}, \citenamefont {Jakuba}, \citenamefont {Kinsey}, \citenamefont
  {McIntyre}, \citenamefont {Sylva},\ and\ \citenamefont
  {Maloney}}]{Camilli2010}%
  \BibitemOpen
  \bibfield  {author} {\bibinfo {author} {\bibfnamefont {R.}~\bibnamefont
  {Camilli}}, \bibinfo {author} {\bibfnamefont {C.~M.}\ \bibnamefont {Reddy}},
  \bibinfo {author} {\bibfnamefont {D.~R.}\ \bibnamefont {Yoerger}}, \bibinfo
  {author} {\bibfnamefont {B.~A.}\ \bibnamefont {{Van Mooy}}}, \bibinfo
  {author} {\bibfnamefont {M.~V.}\ \bibnamefont {Jakuba}}, \bibinfo {author}
  {\bibfnamefont {J.~C.}\ \bibnamefont {Kinsey}}, \bibinfo {author}
  {\bibfnamefont {C.~P.}\ \bibnamefont {McIntyre}}, \bibinfo {author}
  {\bibfnamefont {S.~P.}\ \bibnamefont {Sylva}}, \ and\ \bibinfo {author}
  {\bibfnamefont {J.~V.}\ \bibnamefont {Maloney}},\ }\href {\doibase
  10.1126/science.1195223} {\bibfield  {journal} {\bibinfo  {journal}
  {Science}\ }\textbf {\bibinfo {volume} {330}},\ \bibinfo {pages} {201}
  (\bibinfo {year} {2010})}\BibitemShut {NoStop}%
\bibitem [{\citenamefont {Passow}\ and\ \citenamefont
  {Overton}(2021)}]{Passow2021}%
  \BibitemOpen
  \bibfield  {author} {\bibinfo {author} {\bibfnamefont {U.}~\bibnamefont
  {Passow}}\ and\ \bibinfo {author} {\bibfnamefont {E.~B.}\ \bibnamefont
  {Overton}},\ }\href {\doibase 10.1146/annurev-marine-032320-095153}
  {\bibfield  {journal} {\bibinfo  {journal} {Annual Review of Marine Science}\
  }\textbf {\bibinfo {volume} {13}} (\bibinfo {year} {2021}),\
  10.1146/annurev-marine-032320-095153}\BibitemShut {NoStop}%
\bibitem [{\citenamefont {Li}\ \emph {et~al.}(2015)\citenamefont {Li},
  \citenamefont {Bird}, \citenamefont {Payne}, \citenamefont {Vinhateiro},
  \citenamefont {Kim}, \citenamefont {Davis},\ and\ \citenamefont
  {Loomis}}]{Li2015}%
  \BibitemOpen
  \bibfield  {author} {\bibinfo {author} {\bibfnamefont {Z.}~\bibnamefont
  {Li}}, \bibinfo {author} {\bibfnamefont {A.}~\bibnamefont {Bird}}, \bibinfo
  {author} {\bibfnamefont {J.}~\bibnamefont {Payne}}, \bibinfo {author}
  {\bibfnamefont {N.}~\bibnamefont {Vinhateiro}}, \bibinfo {author}
  {\bibfnamefont {Y.}~\bibnamefont {Kim}}, \bibinfo {author} {\bibfnamefont
  {C.}~\bibnamefont {Davis}}, \ and\ \bibinfo {author} {\bibfnamefont
  {N.}~\bibnamefont {Loomis}},\ }\href
  {https://www.researchgate.net/publication/305500821_Technical_Reports_for_Deepwater_Horizon_Water_Column_Injury_Assessment_Oil_Particle_Data_from_the_Deepwater_Horizon_Oil_Spill}
  {\emph {\bibinfo {title} {{Technical Reports for Deepwater Horizon Water
  Column Injury Assessment: Oil Particle Data from the Deepwater Horizon Oil
  Spill}}}},\ \bibinfo {type} {Tech. Rep.}\ (\bibinfo  {institution} {RPS
  ASA},\ \bibinfo {year} {2015})\BibitemShut {NoStop}%
\bibitem [{\citenamefont {Ryerson}\ \emph {et~al.}(2012)\citenamefont
  {Ryerson}, \citenamefont {Camilli}, \citenamefont {Kessler}, \citenamefont
  {Kujawinski}, \citenamefont {Reddy}, \citenamefont {Valentine}, \citenamefont
  {Atlas}, \citenamefont {Blake}, \citenamefont {{De Gouw}}, \citenamefont
  {Meinardi}, \citenamefont {Parrish}, \citenamefont {Peischl}, \citenamefont
  {Seewald},\ and\ \citenamefont {Warneke}}]{Ryerson2012}%
  \BibitemOpen
  \bibfield  {author} {\bibinfo {author} {\bibfnamefont {T.~B.}\ \bibnamefont
  {Ryerson}}, \bibinfo {author} {\bibfnamefont {R.}~\bibnamefont {Camilli}},
  \bibinfo {author} {\bibfnamefont {J.~D.}\ \bibnamefont {Kessler}}, \bibinfo
  {author} {\bibfnamefont {E.~B.}\ \bibnamefont {Kujawinski}}, \bibinfo
  {author} {\bibfnamefont {C.~M.}\ \bibnamefont {Reddy}}, \bibinfo {author}
  {\bibfnamefont {D.~L.}\ \bibnamefont {Valentine}}, \bibinfo {author}
  {\bibfnamefont {E.}~\bibnamefont {Atlas}}, \bibinfo {author} {\bibfnamefont
  {D.~R.}\ \bibnamefont {Blake}}, \bibinfo {author} {\bibfnamefont
  {J.}~\bibnamefont {{De Gouw}}}, \bibinfo {author} {\bibfnamefont
  {S.}~\bibnamefont {Meinardi}}, \bibinfo {author} {\bibfnamefont {D.~D.}\
  \bibnamefont {Parrish}}, \bibinfo {author} {\bibfnamefont {J.}~\bibnamefont
  {Peischl}}, \bibinfo {author} {\bibfnamefont {J.~S.}\ \bibnamefont
  {Seewald}}, \ and\ \bibinfo {author} {\bibfnamefont {C.}~\bibnamefont
  {Warneke}},\ }\href {\doibase 10.1073/pnas.1110564109} {\bibfield  {journal}
  {\bibinfo  {journal} {Proceedings of the National Academy of Sciences of the
  United States of America}\ }\textbf {\bibinfo {volume} {109}},\ \bibinfo
  {pages} {20246} (\bibinfo {year} {2012})}\BibitemShut {NoStop}%
\bibitem [{\citenamefont {North}\ \emph {et~al.}(2015)\citenamefont {North},
  \citenamefont {Adams}, \citenamefont {Thessen}, \citenamefont {Schlag},
  \citenamefont {He}, \citenamefont {Socolofsky}, \citenamefont {Masutani},\
  and\ \citenamefont {Peckham}}]{North2015}%
  \BibitemOpen
  \bibfield  {author} {\bibinfo {author} {\bibfnamefont {E.~W.}\ \bibnamefont
  {North}}, \bibinfo {author} {\bibfnamefont {E.~E.}\ \bibnamefont {Adams}},
  \bibinfo {author} {\bibfnamefont {A.~E.}\ \bibnamefont {Thessen}}, \bibinfo
  {author} {\bibfnamefont {Z.}~\bibnamefont {Schlag}}, \bibinfo {author}
  {\bibfnamefont {R.}~\bibnamefont {He}}, \bibinfo {author} {\bibfnamefont
  {S.~A.}\ \bibnamefont {Socolofsky}}, \bibinfo {author} {\bibfnamefont
  {S.~M.}\ \bibnamefont {Masutani}}, \ and\ \bibinfo {author} {\bibfnamefont
  {S.~D.}\ \bibnamefont {Peckham}},\ }\href {\doibase
  10.1088/1748-9326/10/2/024016} {\bibfield  {journal} {\bibinfo  {journal}
  {Environmental Research Letters}\ }\textbf {\bibinfo {volume} {10}} (\bibinfo
  {year} {2015}),\ 10.1088/1748-9326/10/2/024016}\BibitemShut {NoStop}%
\bibitem [{\citenamefont {French-McCay}\ \emph {et~al.}(2019)\citenamefont
  {French-McCay}, \citenamefont {Crowley},\ and\ \citenamefont
  {McStay}}]{French-McCay2019}%
  \BibitemOpen
  \bibfield  {author} {\bibinfo {author} {\bibfnamefont {D.}~\bibnamefont
  {French-McCay}}, \bibinfo {author} {\bibfnamefont {D.}~\bibnamefont
  {Crowley}}, \ and\ \bibinfo {author} {\bibfnamefont {L.}~\bibnamefont
  {McStay}},\ }\href {\doibase 10.1016/J.MARPOLBUL.2019.07.038} {\bibfield
  {journal} {\bibinfo  {journal} {Marine Pollution Bulletin}\ }\textbf
  {\bibinfo {volume} {146}},\ \bibinfo {pages} {779} (\bibinfo {year}
  {2019})}\BibitemShut {NoStop}%
\bibitem [{\citenamefont {Passow}\ and\ \citenamefont
  {Hetland}(2016)}]{Passow2016}%
  \BibitemOpen
  \bibfield  {author} {\bibinfo {author} {\bibfnamefont {U.}~\bibnamefont
  {Passow}}\ and\ \bibinfo {author} {\bibfnamefont {R.}~\bibnamefont
  {Hetland}},\ }\href {\doibase 10.5670/oceanog.2016.73} {\bibfield  {journal}
  {\bibinfo  {journal} {Oceanography}\ }\textbf {\bibinfo {volume} {29}},\
  \bibinfo {pages} {88} (\bibinfo {year} {2016})}\BibitemShut {NoStop}%
\bibitem [{\citenamefont {Vilc{\'{a}}ez}\ \emph {et~al.}(2013)\citenamefont
  {Vilc{\'{a}}ez}, \citenamefont {Li},\ and\ \citenamefont
  {Hubbard}}]{Vilcaez2013}%
  \BibitemOpen
  \bibfield  {author} {\bibinfo {author} {\bibfnamefont {J.}~\bibnamefont
  {Vilc{\'{a}}ez}}, \bibinfo {author} {\bibfnamefont {L.}~\bibnamefont {Li}}, \
  and\ \bibinfo {author} {\bibfnamefont {S.~S.}\ \bibnamefont {Hubbard}},\
  }\href {\doibase 10.1186/1467-4866-14-4} {\bibfield  {journal} {\bibinfo
  {journal} {Geochemical Transactions}\ }\textbf {\bibinfo {volume} {14}}
  (\bibinfo {year} {2013}),\ 10.1186/1467-4866-14-4}\BibitemShut {NoStop}%
\bibitem [{\citenamefont {Fernandez}\ \emph {et~al.}(2022)\citenamefont
  {Fernandez}, \citenamefont {Stocker},\ and\ \citenamefont
  {Juarez}}]{Fernandez2022}%
  \BibitemOpen
  \bibfield  {author} {\bibinfo {author} {\bibfnamefont {V.~I.}\ \bibnamefont
  {Fernandez}}, \bibinfo {author} {\bibfnamefont {R.}~\bibnamefont {Stocker}},
  \ and\ \bibinfo {author} {\bibfnamefont {G.}~\bibnamefont {Juarez}},\ }\href
  {\doibase 10.1038/s41598-022-08581-7} {\bibfield  {journal} {\bibinfo
  {journal} {Scientific Reports 2022 12:1}\ }\textbf {\bibinfo {volume} {12}},\
  \bibinfo {pages} {1} (\bibinfo {year} {2022})}\BibitemShut {NoStop}%
\bibitem [{\citenamefont {North}\ \emph {et~al.}(2013)\citenamefont {North},
  \citenamefont {Adams}, \citenamefont {Schlag}, \citenamefont {Sherwood},
  \citenamefont {He}, \citenamefont {Hyun},\ and\ \citenamefont
  {Socolofsky}}]{North2013}%
  \BibitemOpen
  \bibfield  {author} {\bibinfo {author} {\bibfnamefont {E.~W.}\ \bibnamefont
  {North}}, \bibinfo {author} {\bibfnamefont {E.~E.}\ \bibnamefont {Adams}},
  \bibinfo {author} {\bibfnamefont {Z.}~\bibnamefont {Schlag}}, \bibinfo
  {author} {\bibfnamefont {C.~R.}\ \bibnamefont {Sherwood}}, \bibinfo {author}
  {\bibfnamefont {R.}~\bibnamefont {He}}, \bibinfo {author} {\bibfnamefont
  {K.~H.}\ \bibnamefont {Hyun}}, \ and\ \bibinfo {author} {\bibfnamefont
  {S.~A.}\ \bibnamefont {Socolofsky}},\ }in\ \href {\doibase
  10.1029/2011GM001102} {\emph {\bibinfo {booktitle} {Monitoring and Modeling
  the Deepwater Horizon Oil Spill: A Record Breaking Enterprise}}}\ (\bibinfo
  {publisher} {wiley},\ \bibinfo {year} {2013})\ pp.\ \bibinfo {pages}
  {217--226}\BibitemShut {NoStop}%
\bibitem [{\citenamefont {Daly}\ \emph {et~al.}(2016)\citenamefont {Daly},
  \citenamefont {Passow}, \citenamefont {Chanton},\ and\ \citenamefont
  {Hollander}}]{Daly2016}%
  \BibitemOpen
  \bibfield  {author} {\bibinfo {author} {\bibfnamefont {K.~L.}\ \bibnamefont
  {Daly}}, \bibinfo {author} {\bibfnamefont {U.}~\bibnamefont {Passow}},
  \bibinfo {author} {\bibfnamefont {J.}~\bibnamefont {Chanton}}, \ and\
  \bibinfo {author} {\bibfnamefont {D.}~\bibnamefont {Hollander}},\ }\href
  {\doibase 10.1016/j.ancene.2016.01.006} {\bibfield  {journal} {\bibinfo
  {journal} {Anthropocene}\ }\textbf {\bibinfo {volume} {13}},\ \bibinfo
  {pages} {18} (\bibinfo {year} {2016})}\BibitemShut {NoStop}%
\bibitem [{\citenamefont {Socolofsky}\ \emph {et~al.}(2019)\citenamefont
  {Socolofsky}, \citenamefont {Gros}, \citenamefont {North}, \citenamefont
  {Boufadel}, \citenamefont {Parkerton},\ and\ \citenamefont
  {Adams}}]{Socolofsky2019}%
  \BibitemOpen
  \bibfield  {author} {\bibinfo {author} {\bibfnamefont {S.~A.}\ \bibnamefont
  {Socolofsky}}, \bibinfo {author} {\bibfnamefont {J.}~\bibnamefont {Gros}},
  \bibinfo {author} {\bibfnamefont {E.}~\bibnamefont {North}}, \bibinfo
  {author} {\bibfnamefont {M.~C.}\ \bibnamefont {Boufadel}}, \bibinfo {author}
  {\bibfnamefont {T.~F.}\ \bibnamefont {Parkerton}}, \ and\ \bibinfo {author}
  {\bibfnamefont {E.~E.}\ \bibnamefont {Adams}},\ }\href {\doibase
  10.1016/j.marpolbul.2019.04.018} {\bibfield  {journal} {\bibinfo  {journal}
  {Marine Pollution Bulletin}\ }\textbf {\bibinfo {volume} {143}},\ \bibinfo
  {pages} {204} (\bibinfo {year} {2019})}\BibitemShut {NoStop}%
\bibitem [{\citenamefont {Hickl}\ and\ \citenamefont
  {Juarez}(2022{\natexlab{a}})}]{Hickl2022}%
  \BibitemOpen
  \bibfield  {author} {\bibinfo {author} {\bibfnamefont {V.}~\bibnamefont
  {Hickl}}\ and\ \bibinfo {author} {\bibfnamefont {G.}~\bibnamefont {Juarez}},\
  }\href@noop {} {\bibfield  {journal} {\bibinfo  {journal} {Marine Pollution
  Bulletin}\ }\textbf {\bibinfo {volume} {178}},\ \bibinfo {pages} {113645}
  (\bibinfo {year} {2022}{\natexlab{a}})}\BibitemShut {NoStop}%
\bibitem [{\citenamefont {Omarova}\ \emph {et~al.}(2019)\citenamefont
  {Omarova}, \citenamefont {Swientoniewski}, \citenamefont {{Mkam Tsengam}},
  \citenamefont {Blake}, \citenamefont {John}, \citenamefont {McCormick},
  \citenamefont {Bothun}, \citenamefont {Raghavan},\ and\ \citenamefont
  {Bose}}]{Omarova2019}%
  \BibitemOpen
  \bibfield  {author} {\bibinfo {author} {\bibfnamefont {M.}~\bibnamefont
  {Omarova}}, \bibinfo {author} {\bibfnamefont {L.~T.}\ \bibnamefont
  {Swientoniewski}}, \bibinfo {author} {\bibfnamefont {I.~K.}\ \bibnamefont
  {{Mkam Tsengam}}}, \bibinfo {author} {\bibfnamefont {D.~A.}\ \bibnamefont
  {Blake}}, \bibinfo {author} {\bibfnamefont {V.}~\bibnamefont {John}},
  \bibinfo {author} {\bibfnamefont {A.}~\bibnamefont {McCormick}}, \bibinfo
  {author} {\bibfnamefont {G.~D.}\ \bibnamefont {Bothun}}, \bibinfo {author}
  {\bibfnamefont {S.~R.}\ \bibnamefont {Raghavan}}, \ and\ \bibinfo {author}
  {\bibfnamefont {A.}~\bibnamefont {Bose}},\ }\href {\doibase
  10.1021/acssuschemeng.9b01923} {\bibfield  {journal} {\bibinfo  {journal}
  {ACS Sustainable Chemistry \& Engineering}\ ,\ \bibinfo {pages}
  {acssuschemeng.9b01923}} (\bibinfo {year} {2019})}\BibitemShut {NoStop}%
\bibitem [{\citenamefont {Hickl}\ and\ \citenamefont
  {Juarez}(2022{\natexlab{b}})}]{Hickl2022b}%
  \BibitemOpen
  \bibfield  {author} {\bibinfo {author} {\bibfnamefont {V.}~\bibnamefont
  {Hickl}}\ and\ \bibinfo {author} {\bibfnamefont {G.}~\bibnamefont {Juarez}},\
  }\href {\doibase 10.1039/D2SM00813K} {\bibfield  {journal} {\bibinfo
  {journal} {Soft Matter}\ }\textbf {\bibinfo {volume} {18}},\ \bibinfo {pages}
  {7217} (\bibinfo {year} {2022}{\natexlab{b}})}\BibitemShut {NoStop}%
\bibitem [{\citenamefont {White}\ \emph {et~al.}(2020)\citenamefont {White},
  \citenamefont {Jalali}, \citenamefont {Boufadel},\ and\ \citenamefont
  {Sheng}}]{White2020}%
  \BibitemOpen
  \bibfield  {author} {\bibinfo {author} {\bibfnamefont {A.~R.}\ \bibnamefont
  {White}}, \bibinfo {author} {\bibfnamefont {M.}~\bibnamefont {Jalali}},
  \bibinfo {author} {\bibfnamefont {M.~C.}\ \bibnamefont {Boufadel}}, \ and\
  \bibinfo {author} {\bibfnamefont {J.}~\bibnamefont {Sheng}},\ }\href
  {\doibase 10.1038/s41598-020-61214-9} {\bibfield  {journal} {\bibinfo
  {journal} {Scientific Reports 2020 10:1}\ }\textbf {\bibinfo {volume} {10}},\
  \bibinfo {pages} {1} (\bibinfo {year} {2020})}\BibitemShut {NoStop}%
\bibitem [{\citenamefont {Ki{\o}rboe}(2001)}]{Kiorboe2001}%
  \BibitemOpen
  \bibfield  {author} {\bibinfo {author} {\bibfnamefont {T.}~\bibnamefont
  {Ki{\o}rboe}},\ }\href {\doibase 10.3989/SCIMAR.2001.65S257} {\bibfield
  {journal} {\bibinfo  {journal} {Scientia Marina}\ }\textbf {\bibinfo {volume}
  {65}},\ \bibinfo {pages} {57} (\bibinfo {year} {2001})}\BibitemShut {NoStop}%
\bibitem [{\citenamefont {Valentine}\ \emph {et~al.}(2014)\citenamefont
  {Valentine}, \citenamefont {Fisher}, \citenamefont {Bagby}, \citenamefont
  {Nelson}, \citenamefont {Reddy}, \citenamefont {Sylva},\ and\ \citenamefont
  {Wood}}]{Valentine2014}%
  \BibitemOpen
  \bibfield  {author} {\bibinfo {author} {\bibfnamefont {D.~L.}\ \bibnamefont
  {Valentine}}, \bibinfo {author} {\bibfnamefont {G.~B.}\ \bibnamefont
  {Fisher}}, \bibinfo {author} {\bibfnamefont {S.~C.}\ \bibnamefont {Bagby}},
  \bibinfo {author} {\bibfnamefont {R.~K.}\ \bibnamefont {Nelson}}, \bibinfo
  {author} {\bibfnamefont {C.~M.}\ \bibnamefont {Reddy}}, \bibinfo {author}
  {\bibfnamefont {S.~P.}\ \bibnamefont {Sylva}}, \ and\ \bibinfo {author}
  {\bibfnamefont {M.~A.}\ \bibnamefont {Wood}},\ }\href {\doibase
  10.1073/PNAS.1414873111/SUPPL_FILE/PNAS.1414873111.SD01.XLS} {\bibfield
  {journal} {\bibinfo  {journal} {Proceedings of the National Academy of
  Sciences of the United States of America}\ }\textbf {\bibinfo {volume}
  {111}},\ \bibinfo {pages} {15906} (\bibinfo {year} {2014})}\BibitemShut
  {NoStop}%
\bibitem [{\citenamefont {Brakstad}\ \emph {et~al.}(2018)\citenamefont
  {Brakstad}, \citenamefont {Lewis},\ and\ \citenamefont
  {Beegle-Krause}}]{Brakstad2018}%
  \BibitemOpen
  \bibfield  {author} {\bibinfo {author} {\bibfnamefont {O.~G.}\ \bibnamefont
  {Brakstad}}, \bibinfo {author} {\bibfnamefont {A.}~\bibnamefont {Lewis}}, \
  and\ \bibinfo {author} {\bibfnamefont {C.~J.}\ \bibnamefont
  {Beegle-Krause}},\ }\href {\doibase 10.1016/J.MARPOLBUL.2018.07.028}
  {\bibfield  {journal} {\bibinfo  {journal} {Marine Pollution Bulletin}\
  }\textbf {\bibinfo {volume} {135}},\ \bibinfo {pages} {346} (\bibinfo {year}
  {2018})}\BibitemShut {NoStop}%
\bibitem [{\citenamefont {Ross}\ \emph {et~al.}(2021)\citenamefont {Ross},
  \citenamefont {Hollander}, \citenamefont {Saupe}, \citenamefont {Burd},
  \citenamefont {Gilbert},\ and\ \citenamefont {Quigg}}]{Ross2021}%
  \BibitemOpen
  \bibfield  {author} {\bibinfo {author} {\bibfnamefont {J.}~\bibnamefont
  {Ross}}, \bibinfo {author} {\bibfnamefont {D.}~\bibnamefont {Hollander}},
  \bibinfo {author} {\bibfnamefont {S.}~\bibnamefont {Saupe}}, \bibinfo
  {author} {\bibfnamefont {A.~B.}\ \bibnamefont {Burd}}, \bibinfo {author}
  {\bibfnamefont {S.}~\bibnamefont {Gilbert}}, \ and\ \bibinfo {author}
  {\bibfnamefont {A.}~\bibnamefont {Quigg}},\ }\href {\doibase
  10.1016/J.MARPOLBUL.2021.112025} {\bibfield  {journal} {\bibinfo  {journal}
  {Marine Pollution Bulletin}\ }\textbf {\bibinfo {volume} {165}},\ \bibinfo
  {pages} {112025} (\bibinfo {year} {2021})}\BibitemShut {NoStop}%
\bibitem [{\citenamefont {Gregson}\ \emph {et~al.}(2021)\citenamefont
  {Gregson}, \citenamefont {McKew}, \citenamefont {Holland}, \citenamefont
  {Nedwed}, \citenamefont {Prince},\ and\ \citenamefont
  {McGenity}}]{Gregson2021}%
  \BibitemOpen
  \bibfield  {author} {\bibinfo {author} {\bibfnamefont {B.~H.}\ \bibnamefont
  {Gregson}}, \bibinfo {author} {\bibfnamefont {B.~A.}\ \bibnamefont {McKew}},
  \bibinfo {author} {\bibfnamefont {R.~D.}\ \bibnamefont {Holland}}, \bibinfo
  {author} {\bibfnamefont {T.~J.}\ \bibnamefont {Nedwed}}, \bibinfo {author}
  {\bibfnamefont {R.~C.}\ \bibnamefont {Prince}}, \ and\ \bibinfo {author}
  {\bibfnamefont {T.~J.}\ \bibnamefont {McGenity}},\ }\href {\doibase
  10.3389/FMARS.2021.619484/BIBTEX} {\bibfield  {journal} {\bibinfo  {journal}
  {Frontiers in Marine Science}\ }\textbf {\bibinfo {volume} {8}},\ \bibinfo
  {pages} {11} (\bibinfo {year} {2021})}\BibitemShut {NoStop}%
\bibitem [{\citenamefont {S{\l}omka}\ \emph {et~al.}(2020)\citenamefont
  {S{\l}omka}, \citenamefont {Alcolombri}, \citenamefont {Secchi},
  \citenamefont {Stocker},\ and\ \citenamefont {Fernandez}}]{Slomka2020}%
  \BibitemOpen
  \bibfield  {author} {\bibinfo {author} {\bibfnamefont {J.}~\bibnamefont
  {S{\l}omka}}, \bibinfo {author} {\bibfnamefont {U.}~\bibnamefont
  {Alcolombri}}, \bibinfo {author} {\bibfnamefont {E.}~\bibnamefont {Secchi}},
  \bibinfo {author} {\bibfnamefont {R.}~\bibnamefont {Stocker}}, \ and\
  \bibinfo {author} {\bibfnamefont {V.~I.}\ \bibnamefont {Fernandez}},\ }\href
  {\doibase 10.1088/1367-2630/AB73C9} {\bibfield  {journal} {\bibinfo
  {journal} {New Journal of Physics}\ }\textbf {\bibinfo {volume} {22}},\
  \bibinfo {pages} {043016} (\bibinfo {year} {2020})},\ \Eprint
  {http://arxiv.org/abs/1908.08376} {arXiv:1908.08376} \BibitemShut {NoStop}%
\bibitem [{\citenamefont {Abbasi}\ \emph {et~al.}(2018)\citenamefont {Abbasi},
  \citenamefont {Bothun},\ and\ \citenamefont {Bose}}]{Abbasi2018}%
  \BibitemOpen
  \bibfield  {author} {\bibinfo {author} {\bibfnamefont {A.}~\bibnamefont
  {Abbasi}}, \bibinfo {author} {\bibfnamefont {G.~D.}\ \bibnamefont {Bothun}},
  \ and\ \bibinfo {author} {\bibfnamefont {A.}~\bibnamefont {Bose}},\ }\href
  {\doibase 10.1021/acs.langmuir.8b00082} {\bibfield  {journal} {\bibinfo
  {journal} {Langmuir}\ }\textbf {\bibinfo {volume} {34}},\ \bibinfo {pages}
  {5352} (\bibinfo {year} {2018})}\BibitemShut {NoStop}%
\bibitem [{\citenamefont {Krishnamurthy}\ \emph {et~al.}(2020)\citenamefont
  {Krishnamurthy}, \citenamefont {Li}, \citenamefont {{Benoit du Rey}},
  \citenamefont {Cambournac}, \citenamefont {Larson}, \citenamefont {Li},\ and\
  \citenamefont {Prakash}}]{Krishnamurthy}%
  \BibitemOpen
  \bibfield  {author} {\bibinfo {author} {\bibfnamefont {D.}~\bibnamefont
  {Krishnamurthy}}, \bibinfo {author} {\bibfnamefont {H.}~\bibnamefont {Li}},
  \bibinfo {author} {\bibfnamefont {F.}~\bibnamefont {{Benoit du Rey}}},
  \bibinfo {author} {\bibfnamefont {P.}~\bibnamefont {Cambournac}}, \bibinfo
  {author} {\bibfnamefont {A.~G.}\ \bibnamefont {Larson}}, \bibinfo {author}
  {\bibfnamefont {E.}~\bibnamefont {Li}}, \ and\ \bibinfo {author}
  {\bibfnamefont {M.}~\bibnamefont {Prakash}},\ }\href {\doibase
  10.1038/s41592-020-0924-7} {\bibfield  {journal} {\bibinfo  {journal} {Nature
  Methods}\ }\textbf {\bibinfo {volume} {17}},\ \bibinfo {pages} {1040}
  (\bibinfo {year} {2020})}\BibitemShut {NoStop}%
\bibitem [{\citenamefont {Daling}\ \emph {et~al.}(2014)\citenamefont {Daling},
  \citenamefont {Leirvik}, \citenamefont {Alm{\aa}s}, \citenamefont {Brandvik},
  \citenamefont {Hansen}, \citenamefont {Lewis},\ and\ \citenamefont
  {Reed}}]{Daling2014}%
  \BibitemOpen
  \bibfield  {author} {\bibinfo {author} {\bibfnamefont {P.~S.}\ \bibnamefont
  {Daling}}, \bibinfo {author} {\bibfnamefont {F.}~\bibnamefont {Leirvik}},
  \bibinfo {author} {\bibfnamefont {I.~K.}\ \bibnamefont {Alm{\aa}s}}, \bibinfo
  {author} {\bibfnamefont {P.~J.}\ \bibnamefont {Brandvik}}, \bibinfo {author}
  {\bibfnamefont {B.~H.}\ \bibnamefont {Hansen}}, \bibinfo {author}
  {\bibfnamefont {A.}~\bibnamefont {Lewis}}, \ and\ \bibinfo {author}
  {\bibfnamefont {M.}~\bibnamefont {Reed}},\ }\href {\doibase
  10.1016/j.marpolbul.2014.07.005} {\bibfield  {journal} {\bibinfo  {journal}
  {Marine Pollution Bulletin}\ }\textbf {\bibinfo {volume} {87}},\ \bibinfo
  {pages} {300} (\bibinfo {year} {2014})}\BibitemShut {NoStop}%
\bibitem [{\citenamefont {Hazen}\ \emph {et~al.}(2010)\citenamefont {Hazen},
  \citenamefont {Dubinsky}, \citenamefont {DeSantis}, \citenamefont {Andersen},
  \citenamefont {Piceno}, \citenamefont {Singh}, \citenamefont {Jansson},
  \citenamefont {Probst}, \citenamefont {Borglin}, \citenamefont {Fortney},
  \citenamefont {Stringfellow}, \citenamefont {Bill}, \citenamefont {Conrad},
  \citenamefont {Tom}, \citenamefont {Chavarria}, \citenamefont {Alusi},
  \citenamefont {Lamendella}, \citenamefont {Joyner}, \citenamefont {Spier},
  \citenamefont {Baelum}, \citenamefont {Auer}, \citenamefont {Zemla},
  \citenamefont {Chakraborty}, \citenamefont {Sonnenthal}, \citenamefont
  {D'haeseleer}, \citenamefont {Holman}, \citenamefont {Osman}, \citenamefont
  {Lu}, \citenamefont {{Van Nostrand}}, \citenamefont {Deng}, \citenamefont
  {Zhou},\ and\ \citenamefont {Mason}}]{Hazen2010}%
  \BibitemOpen
  \bibfield  {author} {\bibinfo {author} {\bibfnamefont {T.~C.}\ \bibnamefont
  {Hazen}}, \bibinfo {author} {\bibfnamefont {E.~A.}\ \bibnamefont {Dubinsky}},
  \bibinfo {author} {\bibfnamefont {T.~Z.}\ \bibnamefont {DeSantis}}, \bibinfo
  {author} {\bibfnamefont {G.~L.}\ \bibnamefont {Andersen}}, \bibinfo {author}
  {\bibfnamefont {Y.~M.}\ \bibnamefont {Piceno}}, \bibinfo {author}
  {\bibfnamefont {N.}~\bibnamefont {Singh}}, \bibinfo {author} {\bibfnamefont
  {J.~K.}\ \bibnamefont {Jansson}}, \bibinfo {author} {\bibfnamefont
  {A.}~\bibnamefont {Probst}}, \bibinfo {author} {\bibfnamefont {S.~E.}\
  \bibnamefont {Borglin}}, \bibinfo {author} {\bibfnamefont {J.~L.}\
  \bibnamefont {Fortney}}, \bibinfo {author} {\bibfnamefont {W.~T.}\
  \bibnamefont {Stringfellow}}, \bibinfo {author} {\bibfnamefont
  {M.}~\bibnamefont {Bill}}, \bibinfo {author} {\bibfnamefont {M.~E.}\
  \bibnamefont {Conrad}}, \bibinfo {author} {\bibfnamefont {L.~M.}\
  \bibnamefont {Tom}}, \bibinfo {author} {\bibfnamefont {K.~L.}\ \bibnamefont
  {Chavarria}}, \bibinfo {author} {\bibfnamefont {T.~R.}\ \bibnamefont
  {Alusi}}, \bibinfo {author} {\bibfnamefont {R.}~\bibnamefont {Lamendella}},
  \bibinfo {author} {\bibfnamefont {D.~C.}\ \bibnamefont {Joyner}}, \bibinfo
  {author} {\bibfnamefont {C.}~\bibnamefont {Spier}}, \bibinfo {author}
  {\bibfnamefont {J.}~\bibnamefont {Baelum}}, \bibinfo {author} {\bibfnamefont
  {M.}~\bibnamefont {Auer}}, \bibinfo {author} {\bibfnamefont {M.~L.}\
  \bibnamefont {Zemla}}, \bibinfo {author} {\bibfnamefont {R.}~\bibnamefont
  {Chakraborty}}, \bibinfo {author} {\bibfnamefont {E.~L.}\ \bibnamefont
  {Sonnenthal}}, \bibinfo {author} {\bibfnamefont {P.}~\bibnamefont
  {D'haeseleer}}, \bibinfo {author} {\bibfnamefont {H.-Y.~N.}\ \bibnamefont
  {Holman}}, \bibinfo {author} {\bibfnamefont {S.}~\bibnamefont {Osman}},
  \bibinfo {author} {\bibfnamefont {Z.}~\bibnamefont {Lu}}, \bibinfo {author}
  {\bibfnamefont {J.~D.}\ \bibnamefont {{Van Nostrand}}}, \bibinfo {author}
  {\bibfnamefont {Y.}~\bibnamefont {Deng}}, \bibinfo {author} {\bibfnamefont
  {J.}~\bibnamefont {Zhou}}, \ and\ \bibinfo {author} {\bibfnamefont {O.~U.}\
  \bibnamefont {Mason}},\ }\href {\doibase 10.1126/science.1195979} {\bibfield
  {journal} {\bibinfo  {journal} {Science (New York, N.Y.)}\ }\textbf {\bibinfo
  {volume} {330}},\ \bibinfo {pages} {204} (\bibinfo {year}
  {2010})}\BibitemShut {NoStop}%
\bibitem [{\citenamefont {Bochdansky}\ \emph {et~al.}(2016)\citenamefont
  {Bochdansky}, \citenamefont {Clouse},\ and\ \citenamefont
  {Herndl}}]{Bochdansky2016}%
  \BibitemOpen
  \bibfield  {author} {\bibinfo {author} {\bibfnamefont {A.~B.}\ \bibnamefont
  {Bochdansky}}, \bibinfo {author} {\bibfnamefont {M.~A.}\ \bibnamefont
  {Clouse}}, \ and\ \bibinfo {author} {\bibfnamefont {G.~J.}\ \bibnamefont
  {Herndl}},\ }\href {\doibase 10.1038/srep22633} {\bibfield  {journal}
  {\bibinfo  {journal} {Scientific Reports 2016 6:1}\ }\textbf {\bibinfo
  {volume} {6}},\ \bibinfo {pages} {1} (\bibinfo {year} {2016})}\BibitemShut
  {NoStop}%
\bibitem [{\citenamefont {Taute}(2020)}]{Taute2020}%
  \BibitemOpen
  \bibfield  {author} {\bibinfo {author} {\bibfnamefont {K.~M.}\ \bibnamefont
  {Taute}},\ }\href {\doibase 10.1038/s41592-020-0939-0} {\bibfield  {journal}
  {\bibinfo  {journal} {Nature Methods 2020 17:10}\ }\textbf {\bibinfo {volume}
  {17}},\ \bibinfo {pages} {965} (\bibinfo {year} {2020})}\BibitemShut
  {NoStop}%
\bibitem [{\citenamefont {Blasco}(1978)}]{Blasco1978}%
  \BibitemOpen
  \bibfield  {author} {\bibinfo {author} {\bibfnamefont {D.}~\bibnamefont
  {Blasco}},\ }\href {\doibase 10.1007/BF00393819} {\bibfield  {journal}
  {\bibinfo  {journal} {Marine Biology 1978 46:1}\ }\textbf {\bibinfo {volume}
  {46}},\ \bibinfo {pages} {41} (\bibinfo {year} {1978})}\BibitemShut {NoStop}%
\bibitem [{\citenamefont {DeVries}(2022)}]{DeVries2022}%
  \BibitemOpen
  \bibfield  {author} {\bibinfo {author} {\bibfnamefont {T.}~\bibnamefont
  {DeVries}},\ }\href {\doibase 10.1146/ANNUREV-ENVIRON-120920-111307}
  {\bibfield  {journal} {\bibinfo  {journal} {Annual Review of Environment and
  Resources}\ }\textbf {\bibinfo {volume} {47}},\ \bibinfo {pages} {317}
  (\bibinfo {year} {2022})}\BibitemShut {NoStop}%
\bibitem [{\citenamefont {Moran}\ \emph {et~al.}(2022)\citenamefont {Moran},
  \citenamefont {Kujawinski}, \citenamefont {Schroer}, \citenamefont {Amin},
  \citenamefont {Bates}, \citenamefont {Bertrand}, \citenamefont {Braakman},
  \citenamefont {Brown}, \citenamefont {Covert}, \citenamefont {Doney},
  \citenamefont {Dyhrman}, \citenamefont {Edison}, \citenamefont {Eren},
  \citenamefont {Levine}, \citenamefont {Li}, \citenamefont {Ross},
  \citenamefont {Saito}, \citenamefont {Santoro}, \citenamefont {Segr{\`{e}}},
  \citenamefont {Shade}, \citenamefont {Sullivan},\ and\ \citenamefont
  {Vardi}}]{Moran2022}%
  \BibitemOpen
  \bibfield  {author} {\bibinfo {author} {\bibfnamefont {M.~A.}\ \bibnamefont
  {Moran}}, \bibinfo {author} {\bibfnamefont {E.~B.}\ \bibnamefont
  {Kujawinski}}, \bibinfo {author} {\bibfnamefont {W.~F.}\ \bibnamefont
  {Schroer}}, \bibinfo {author} {\bibfnamefont {S.~A.}\ \bibnamefont {Amin}},
  \bibinfo {author} {\bibfnamefont {N.~R.}\ \bibnamefont {Bates}}, \bibinfo
  {author} {\bibfnamefont {E.~M.}\ \bibnamefont {Bertrand}}, \bibinfo {author}
  {\bibfnamefont {R.}~\bibnamefont {Braakman}}, \bibinfo {author}
  {\bibfnamefont {C.~T.}\ \bibnamefont {Brown}}, \bibinfo {author}
  {\bibfnamefont {M.~W.}\ \bibnamefont {Covert}}, \bibinfo {author}
  {\bibfnamefont {S.~C.}\ \bibnamefont {Doney}}, \bibinfo {author}
  {\bibfnamefont {S.~T.}\ \bibnamefont {Dyhrman}}, \bibinfo {author}
  {\bibfnamefont {A.~S.}\ \bibnamefont {Edison}}, \bibinfo {author}
  {\bibfnamefont {A.~M.}\ \bibnamefont {Eren}}, \bibinfo {author}
  {\bibfnamefont {N.~M.}\ \bibnamefont {Levine}}, \bibinfo {author}
  {\bibfnamefont {L.}~\bibnamefont {Li}}, \bibinfo {author} {\bibfnamefont
  {A.~C.}\ \bibnamefont {Ross}}, \bibinfo {author} {\bibfnamefont {M.~A.}\
  \bibnamefont {Saito}}, \bibinfo {author} {\bibfnamefont {A.~E.}\ \bibnamefont
  {Santoro}}, \bibinfo {author} {\bibfnamefont {D.}~\bibnamefont
  {Segr{\`{e}}}}, \bibinfo {author} {\bibfnamefont {A.}~\bibnamefont {Shade}},
  \bibinfo {author} {\bibfnamefont {M.~B.}\ \bibnamefont {Sullivan}}, \ and\
  \bibinfo {author} {\bibfnamefont {A.}~\bibnamefont {Vardi}},\ }\href
  {\doibase 10.1038/s41564-022-01090-3} {\bibfield  {journal} {\bibinfo
  {journal} {Nature Microbiology 2022 7:4}\ }\textbf {\bibinfo {volume} {7}},\
  \bibinfo {pages} {508} (\bibinfo {year} {2022})}\BibitemShut {NoStop}%
\bibitem [{\citenamefont {Datta}\ \emph {et~al.}(2016)\citenamefont {Datta},
  \citenamefont {Sliwerska}, \citenamefont {Gore}, \citenamefont {Polz},\ and\
  \citenamefont {Cordero}}]{Datta2016}%
  \BibitemOpen
  \bibfield  {author} {\bibinfo {author} {\bibfnamefont {M.~S.}\ \bibnamefont
  {Datta}}, \bibinfo {author} {\bibfnamefont {E.}~\bibnamefont {Sliwerska}},
  \bibinfo {author} {\bibfnamefont {J.}~\bibnamefont {Gore}}, \bibinfo {author}
  {\bibfnamefont {M.~F.}\ \bibnamefont {Polz}}, \ and\ \bibinfo {author}
  {\bibfnamefont {O.~X.}\ \bibnamefont {Cordero}},\ }\href {\doibase
  10.1038/ncomms11965} {\bibfield  {journal} {\bibinfo  {journal} {Nature
  Communications 2016 7:1}\ }\textbf {\bibinfo {volume} {7}},\ \bibinfo {pages}
  {1} (\bibinfo {year} {2016})}\BibitemShut {NoStop}%
\bibitem [{\citenamefont {Nguyen}\ \emph {et~al.}(2022)\citenamefont {Nguyen},
  \citenamefont {Zakem}, \citenamefont {Ebrahimi}, \citenamefont {Schwartzman},
  \citenamefont {Caglar}, \citenamefont {Amarnath}, \citenamefont {Alcolombri},
  \citenamefont {Peaudecerf}, \citenamefont {Hwa}, \citenamefont {Stocker},
  \citenamefont {Cordero},\ and\ \citenamefont {Levine}}]{Nguyen2022}%
  \BibitemOpen
  \bibfield  {author} {\bibinfo {author} {\bibfnamefont {T.~T.}\ \bibnamefont
  {Nguyen}}, \bibinfo {author} {\bibfnamefont {E.~J.}\ \bibnamefont {Zakem}},
  \bibinfo {author} {\bibfnamefont {A.}~\bibnamefont {Ebrahimi}}, \bibinfo
  {author} {\bibfnamefont {J.}~\bibnamefont {Schwartzman}}, \bibinfo {author}
  {\bibfnamefont {T.}~\bibnamefont {Caglar}}, \bibinfo {author} {\bibfnamefont
  {K.}~\bibnamefont {Amarnath}}, \bibinfo {author} {\bibfnamefont
  {U.}~\bibnamefont {Alcolombri}}, \bibinfo {author} {\bibfnamefont {F.~J.}\
  \bibnamefont {Peaudecerf}}, \bibinfo {author} {\bibfnamefont
  {T.}~\bibnamefont {Hwa}}, \bibinfo {author} {\bibfnamefont {R.}~\bibnamefont
  {Stocker}}, \bibinfo {author} {\bibfnamefont {O.~X.}\ \bibnamefont
  {Cordero}}, \ and\ \bibinfo {author} {\bibfnamefont {N.~M.}\ \bibnamefont
  {Levine}},\ }\href {\doibase 10.1038/s41467-022-29297-2} {\bibfield
  {journal} {\bibinfo  {journal} {Nature Communications 2022 13:1}\ }\textbf
  {\bibinfo {volume} {13}},\ \bibinfo {pages} {1} (\bibinfo {year}
  {2022})}\BibitemShut {NoStop}%
\end{thebibliography}%

\end{document}